\documentstyle[pre,aps,multicol]{revtex}
\renewcommand{\narrowtext}{\begin{multicols}{2} \global\columnwidth20.5pc}
\renewcommand{\widetext}{\end{multicols} \global\columnwidth42.5pc}
\newcommand{\Lrule}{\vspace*{-0.2in}\noindent\vrule width3.5in height.2pt
  depth.2pt \vrule depth0em height1em}
\newcommand{\Rrule}{\vspace{-0.1in}\hfill\vrule depth1em height0pt \vrule
  width3.5in height.2pt depth.2pt\vspace*{-1pc}}

\input{epsf}

\begin{document}
\preprint{cond-mat/9711152, UT-KOMABA/97-12}
\title{
Field theoretic approach 
to the counting problem of Hamiltonian cycles of graphs
}
\author{Saburo Higuchi\cite{hig_email}}
\address{
 Department of Pure and Applied Sciences,\\
 The University of Tokyo \\
  Komaba, Meguro, Tokyo 153-8902, Japan
}
\date{
  November 14, 1997, 
  cond-mat/9711152, 
  UT-KOMABA/97-12, 
  Revised on February 3, 1998}
\maketitle

\begin{abstract}
  A field theoretic representation for the number of Hamiltonian
  cycles of graphs is studied. By integrating out quadratic
  fluctuations around the saddle point, one obtains an estimate for
  the number which reflects characteristics of graphs well.  The
  accuracy of the estimate is verified by applying it to 2d square
  lattices with various boundary conditions.  This is the first
  example of extracting meaningful information from the quadratic
  approximation to the field theory representation.
\end{abstract}

\narrowtext

\section{Introduction}
\begin{figure}
\epsfxsize=170pt
  \begin{center}
    \leavevmode
\epsfbox{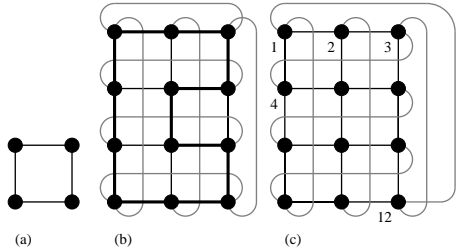}    
  \end{center}
\caption{
Examples of graphs and Hamiltonian cycles on them.
One does not distinguish the base points and the directions of a cycle.
(a) $H(G)=1$ for this graph.
(b) $P(3,4)$, the 2d square lattice with the periodic boundary
condition. A Hamiltonian cycle is drawn in the thick line. 
For this graph, $(q,N)=(4,12)$.
(c) $SP(3,4)$, the 2d square lattice with the skew-periodic boundary
in the horizontal direction. It is locally isomorphic to $P(3,4)$ but
the boundary condition makes it distinct globally.
\label{graphs}
}
\end{figure}

Let $G=(V,E)$ be a graph with  the set of vertices $V=\{r_j\}$ and of the
edges $E=\{e_k\}$. 
A Hamiltonian cycle of a graph is a closed path
which visits each of the vertices once and only once.
I denote the number of all the Hamiltonian cycles of a graph $G$ by ${ H}(G)$: 
\begin{equation}
  { H}(G) = \sum_{\text{\ Hamiltonian cycle on\ } G} 1.
  \label{number_hamiltonian_cycles}
\end{equation}
See Fig.\ref{graphs} for examples.

Hamiltonian cycles have often been used to
model collapsed polymer globules\cite{ClJa:polymerE}. 
The quantity ${ H}(G)$ corresponds to the entropy of a polymer system
in $G$ in a collapsed but disordered phase. 
One can model even more  realistic polymers
by introducing a weight that  depends on the shape of cycles
in (\ref{number_hamiltonian_cycles}).
The polymer melting problem is studied by 
taking into account the bending energy 
with  a weight which depends on the number of turns on the
cycle\cite{BaGaOr:polymermelting}. 
In ref.\cite{DoGaOr:proteinfolding}, the protein
folding problem is studied by incorporating the Van der Waals
potential as well as the bending energy 
in the model (\ref{number_hamiltonian_cycles}).

For homogeneous graphs (lattices) with $N$ vertices,
one expects that $H(G)$ behaves like
\begin{equation}
  H(G) \rightarrow C(G) N^{\gamma-1} \omega^N 
\quad\quad (N\rightarrow\infty),
\label{asymptotic}
\end{equation}
where $\omega$ is defined by
\begin{equation}
  \log \omega=\lim_{N\rightarrow\infty} \frac{1}{N}\log H(G).
\end{equation}
The quantity $\omega$ is supposed to be a universal bulk quantity 
whereas $C(G)$ and
$\gamma$ depend on the detail of graphs \textsl{e.g.} boundary
conditions \cite{ScHiKl:compact}.

A field theory representation for (\ref{number_hamiltonian_cycles})
for arbitrary graphs is introduced  in ref.\cite{OrItDo:hamiltonian} and 
has been used to
study the extended models \cite{BaGaOr:polymermelting,DoGaOr:proteinfolding}.
For homogeneous graphs with the number of vertices $N$ and the
coordination number $q$, 
the saddle point approximation to the representation yields
\begin{equation}
  { H}(G) \simeq  \left(\frac{q}{e}\right)^N
\label{mean_field_saddle_point_estimate}
\end{equation}
or $\omega\simeq q/e$.
This approximation has been proved to be good in many
examples\cite{Suzuki:regular}. 
For a square lattice, $q/e=4/e=1.4715\ldots$ is quite near to the 
exact value $\omega\simeq 1.473$ estimated by the direct
enumeration\cite{CaTh:minimum} and other
methods\cite{ScHiKl:compact,BaBlNiYu:packedloop,DuSa:exact}. 

In the saddle point approximation, however,
graphs with identical $(q,N)$ are not distinguished.
Indeed there is a variety of graphs which has  $(q,N)$ in common. 
Given a graph with a pair $(q,N)$, it is often possible to 
change its boundary conditions to modify the `topology' and the
`moduli' of the graph keeping the pair. Fig. \ref{graphs} shows examples.
Moreover, there are graphs which have identical $(q,N)$ but
have distinct local structures, {\sl e.g.\ } 
the 2d triangular lattice and the 3d cubic lattice.

In this article, I go beyond the saddle point approximation.
I work out the quadratic approximation and find an estimate for $H(G)$ 
whose $G$-dependence is not merely through $(q,N)$ but is more sensible.
To demonstrate the validity of the approximation, 
I examine  2d square lattices with  a variety of boundary conditions
and aspect ratios. 
It is also examined if the estimate $\omega\simeq q/e$
is improved.

\section{Field theoretic representation}
The problem of calculating ${ H}(G)$ is mapped into one in a
lattice field theory on $G$\cite{OrItDo:hamiltonian}. 
This is done by introducing an O($n$) lattice field 
$\vec{\phi}(r)= (\phi_1(r), \ldots, \phi_n(r))$ 
living on $V=\{r_j\}$ with an action
\begin{equation}
S[\vec{\phi}(r)]=\frac12\sum_{
r,r'\in V, 1\le j \le n
} 
         \phi_j(r) (i\Delta^{-1}+\epsilon)_{rr'}\phi_j(r').
\label{action}
\end{equation}
The $N\times N$ matrix $\Delta$ is the adjacency matrix\cite{invertibility}
of the graph $G$:
\begin{equation}
  \Delta_{rr'} =
\left\{
  \begin{array}{ll}
    1 & \text{if $r,r'\in V$ is connected by an $e\in E$}\\
    0 & \text{otherwise}
  \end{array}
\right.
\end{equation}
and an infinitesimal parameter $\epsilon>0$ is introduced for convergence.

The integer ${ H}(G)$ is related to a $2N$-point function by
\begin{eqnarray}
  { H}(G)   
&=&
\lim_{n\rightarrow0}\lim_{\epsilon\rightarrow+0} 
\frac1n\left|\frac{Z_1}{Z_0}\right|,
\label{field_theory_rep}\\
Z_0 &=& \int D\vec{\phi} \quad {\rm e}^{-S[\vec{\phi}(r)]},\\
Z_1 &=& \int D\vec{\phi} \quad {\rm e}^{-S[\vec{\phi}(r)]}
 \prod_{r\in V}\frac{\vec{\phi}^2(r)}{2},
\end{eqnarray}
where  $D\vec{\phi}=\prod_{r\in V, 1\le j\le n} d\phi_j(r)$.
Eq. (\ref{field_theory_rep}) holds 
for arbitrary $G$ with $N>2$
because each terms in the
diagrammatic expansion corresponds to a Hamiltonian cycle.

\section{Approximation}
First I evaluate (\ref{field_theory_rep}) by the saddle point method.
I concentrate on $Z_1$ since $Z_0\rightarrow1$ as $n\rightarrow0$.
When a graph $G$ is homogeneous,  there are  mean field saddle points
which are degenerate on 
\begin{equation}
\{\vec{\phi} |\ \vec{\phi}^2(r)\equiv 
-2iq_\epsilon \}
\simeq \frac{{\rm{O}}(n)}{{\rm{O}}(n-1)},
\end{equation}
where
\begin{equation}
\frac{1}{q_\epsilon}:=\frac{1}{q} -i\epsilon.
\end{equation}
This yields the estimate (\ref{mean_field_saddle_point_estimate}).

Then I consider fluctuations around the saddle point.
It can easily be seen that there are  zero modes corresponding to 
the global O($n)/$O($n-1$) symmetry and sublattice
symmetries\cite{OrItDo:hamiltonian}. 
Therefore I am led to introduce a gauge fix condition
\begin{equation}
  \sum_{r\in V} \phi_j(r) =0  
\quad\quad (2\le j\le n).
\label{gauge_condition}
\end{equation}
By the standard Fadeev-Popov method, I have 
\widetext
\Lrule
\begin{equation}
  Z_1 = \int D\phi \; 
e^{-\frac12 \sum_{r,r'\in V,1\le j\le n} \phi_j(r)(i\Delta^{-1} + \epsilon)_{rr'}\phi_j(r')}
\left| \sum_{r\in V} \phi_1(r)\right|^{n-1}
\frac{\pi^{\frac{n}{2}}}{\Gamma(\frac{n}{2})}
\prod_{2\le j\le n}
\delta\left( \sum_{r\in V}\phi_j(r) \right)
\prod_{r\in V} \frac{\vec{\phi}^2(r)}{2}.
\end{equation}
\Rrule
\narrowtext
The factor 
$
| \sum_{r} \phi_1(r)|^{n-1}\pi^{\frac{n}{2}}/\Gamma(\frac{n}{2})
$
is the Fadeev-Popov determinant multiplied by the volume of the
gauge orbit. 
Actually, this is
nothing but the jacobian for the $n$-dimensional radial coordinate for 
constant modes. 

Expanding the field $\vec{\phi}$ around the saddle point as
\begin{equation}
  \phi_j(r) = 
  \sqrt{-2iq_\epsilon I_{N,n}}\delta_{j1}+ \psi_j(r),
\end{equation}
where
\begin{equation}
  I_{N,n}=1+\frac{n-1}{2N},
\end{equation}
one obtains
\widetext
\Lrule
\begin{eqnarray}
Z_1
&=& 
\left(\frac{q_\epsilon I_{N,n}}{i e}\right)^{N I_{N,n}}
2^{\frac{n-1}{2}} N^{n-1}
\frac{\pi^{\frac{n}{2}}}{\Gamma(\frac{n}{2})}\nonumber \\
& & \times \int D \vec{\psi}\ 
e^{-\frac12 \sum_{r,r'\in V} \psi_1(r) A^{\text{L}}_{rr'}(n) \psi_1(r')}
\times e^{-\frac12 \sum_{r,r'\in V,  2\le j \le n}
                        \psi_j(r) A^{\text{T}}_{rr'}(n) \psi_j(r')}
\times e^{-V_{\rm{int}}(\vec{\psi})}
\times \prod_{2\le j\le n}
\delta\left( \sum_{r\in V}\psi_j(r) \right),
\end{eqnarray}
where $A^{\text{L}}(n)$ and $A^{\text{T}}(n)$ are the inverse propagators of
the longitudinal  mode ($j=1$) and the transverse  modes ($2 \le j \le n$),
respectively:
\begin{eqnarray}
  A^{\text{L}}_{rr'}(n) &=& i(\Delta^{-1})_{rr'} + 
                     \frac{i}{q}I_{N,n}^{-1} \delta_{rr'}
          + (1+I_{N,n}^{-1})\epsilon \delta_{rr'}
   + \frac1N\frac{i}{q_\epsilon}  (1-I_{N,n}^{-1}), 
\label{longitudinal-prop}\\
  A^{\text{T}}_{rr'}(n) &=& i(\Delta^{-1})_{rr'} - 
\frac{i}{q}I_{N,n}^{-1} \delta_{rr'}+(1-I_{N,n}^{-1})\epsilon\delta_{rr'}.
\label{transverse-prop}
\end{eqnarray}
One neglects the interaction terms $V_{\rm{int}}(\vec{\psi})$ and
performs the gaussian integrations to obtain
\begin{equation}
  Z_1 \simeq 
\left(\frac{q I_{N,n}}{i e}\right)^{N I_{N,n}}
(2N)^{\frac{n-1}{2}} 
\frac{\pi^{\frac{n}{2}}}{\Gamma(\frac{n}{2})} 
\left[\frac{(2\pi)^{\frac{N-1}{2}}}{\det'^{\frac{1}{2}} A^{\rm{T}}(n)}\right]^{n-1}
\frac{(2\pi)^{\frac{N}{2}}}{\det^{\frac{1}{2}} A^{\rm{L}}(n)}
\times 2,
\label{z1_finite_n}
\end{equation}
\Rrule
\narrowtext
\noindent
where  $\epsilon$ is sent to $+0$.
The last factor two corresponds to the residual symmetry
$\sum_r\phi_1(r)\leftrightarrow -\sum_r\phi_1(r)$.
The prime on $\det$ means the omission of the eigenvalue for the
constant mode.

One notices that the signature of the real part (the last term) of 
(\ref{transverse-prop}) changes at $n=1$. Thus 
one has to assume $n>1$ to derive (\ref{z1_finite_n}).
Then in eq.(\ref{z1_finite_n}) one  takes the limit $n\rightarrow0$
to obtain the final result
\begin{equation}
H(G)\simeq \!\!
\left(\frac{q}{e}\right)^N \!\!\!\!\!
e^{\frac{1}{2}} (I_{N,0})^{NI_{N,0}}
\sqrt{\frac{\pi}{N}}
\frac{\det'^{\frac{1}{2}}  (A^{\rm{T}}(0)\Delta)}{\det^{\frac{1}{2}} (A^{\rm{L}}(0)\Delta)}.
\label{fluctuation_estimate}
\end{equation}
Note that $\det\Delta\det'A^{\rm{T}}(0)=q\det'(A^{\rm{T}}(0)\Delta). $
Eq. (\ref{fluctuation_estimate}) is nothing but the estimate I desired 
to have. The saddle point result $(q/e)^N$ is
corrected by the ratio of determinants which contains information of
details of the structure of $G$.

In ref.\cite{OrItDo:hamiltonian}, 
it is claimed that the quadratic correction to 
(\ref{mean_field_saddle_point_estimate}) vanishes.
In the present analysis, the Fadeev-Popov method is worked out
to find $\sqrt{\pi/N}$
missing in ref.\cite{OrItDo:hamiltonian}.
As shown in Sec.\ref{square}, the ratio of determinants is not equal to unity
and contributes to $\gamma$ and $C(G)$ non-trivially.
Moreover, inclusion of  $i=\sqrt{-1}$ and $\epsilon>0$ in the action 
(\ref{action}) in the present analysis
enables one to discuss the limit of application of the quadratic
approximation in Sec.\ref{square}.

\section{Square lattices}
\label{square}
To see how eq.(\ref{fluctuation_estimate}) works,
I study concrete examples $P(L_1,L_2)$ and $SP(L_1,L_2)$ shown in
Fig.\ref{graphs}. 
Both are two-dimensional square lattices with the edge lengths
$L_1$ and $L_2$.
The difference between $P$ and $SP$ lies in boundary conditions.
For $P(L_1,L_2)$, the periodic boundary condition is imposed for both two 
directions. 
For $SP(L_1,L_2)$, that across the edge $L_2$ is replace by
the skew-periodic one.
They are good examples to test eq.(\ref{fluctuation_estimate}).
One can switch the boundary condition or vary the aspect ratio
$L_2/L_1$ to make the graph globally distinct while keeping
$(q,N)=(4,L_1\times L_2)$.
The saddle point approximation cannot see the difference among them
but eq.(\ref{fluctuation_estimate}) has a chance to distinguish them.

Graphs $P(L_1,L_2)$ and $SP(L_1,L_2)$ 
can be viewed as discrete tori with different moduli parameters.
Thus it is interesting also in its own right to determine the asymptotic
behaviors of ${ H}(P(L_1,L_2))$ and ${ H}(SP(L_1,L_2))$ in the limit
$L_1,L_2\rightarrow\infty$. 

I analytically evaluate (\ref{fluctuation_estimate})
for $P(L_1,L_2)$  and $SP(L_1,L_2)$.
In the momentum representation,
the determinants become ($\sigma=0$ for $P$ and $\sigma=1$ for $SP$)
\widetext
\Lrule
\begin{eqnarray}
  \det{}' (A^{\rm{T}}(n)\Delta  )
&=& \phantom{2\times}\; {\prod}_{0\le n_j \le L_j-1}'
\left[
1-\frac{1}{2I_{N,n}}
\left(\cos\left(k_1 + \sigma\frac{k_2}{L_1}\right)+\cos k_2\right)\right],\\
  \det (A^{\rm{L}}(n) \Delta )
&=&  2\times{\prod}_{0 \le n_j \le L_j-1}'
\left[
1+\frac{1}{2I_{N,n}}
\left(\cos\left(k_1 + \sigma\frac{k_2}{L_1}\right)+\cos k_2\right)
\right].
\end{eqnarray}
The constant mode $k_1=k_2=0$ is excluded in $\prod'$.
Hereafter, the indices $n_j$ and $k_j$ should be related by $k_j=2\pi n_j/L_j$.

The formula 
\begin{equation}
  \prod_{0\le m \le L-1} 
\left[ x^2 - 2 x  \cos\left(\theta + \frac{2m\pi}{L}\right) + 1 \right]
 = x^{2L} - 2 x^L  \cos(L\theta) +1 
\label{productcosine}
\end{equation}
\Rrule
\narrowtext
\noindent
enables one to obtain
\begin{equation}
\frac{\det'(A^{\rm{T}}(0)\Delta )}{\det(A^{\rm{L}}(0)\Delta )}
= -2 L_1 L_2 I_{L_1L_2,0}\prod_{n_2=0}^{L_2-1}
\frac{u_-(k_2)}{u_+(k_2+\pi)},
\end{equation}
where
\begin{eqnarray}
  u_\pm(k)&=& \left\{
  \begin{array}{lll}
y(k)^{L_1} + y(k)^{-L_1} \pm 2, & \text{for} &P,\\
y(k)^{L_1} + y(k)^{-L_1} - 2\cos k & \text{for} &SP,\\
  \end{array}
\right.\\
 y(k)&=&2I_{L_1L_2,0}- \cos k + 
\sqrt{(2I_{L_1L_2,0} - \cos k)^2-1}.
\end{eqnarray}
Now I specialize to the case where both $L_1$ and $L_2$ are odd.
In the limit $N=L_1 \times L_2 \rightarrow \infty$ with
$R=L_2/L_1$ fixed, 
it is allowed to approximate $\prod_{n_2}$ by $\exp(\int dk_2\; \log)$
after separating quickly oscillating factors.
In this limit, assuming $0<R\le1$ for periodic case, I obtain
\begin{eqnarray}
H(P(L_1,L_2))\!\!
&\simeq&\!\!\left(\frac{q}{e}\right)^N \!\!\!\!\!
\sqrt{\frac{\pi}{2}}
\frac{
\sin(\frac{1}{2R})^{\frac12}
\exp(\frac{9\pi^2}{16R^2}-\frac{1}{2R})^{\frac12}}
{\cosh^2(\frac{\pi^2}{4R^2}-\frac{1}{2R})^{\frac12}}
\label{oop-analytic}\\
H(SP(L_1,L_2))\!\!&\simeq&\!\!
\left(\frac{q}{e}\right)^N\!\!\!\!\!
\sqrt{2\pi}
\frac{
\sin(\frac{1}{2R})^{\frac12}
\left|\sinh(\frac{9\pi^2}{16R^2}-\frac{1}{2R})^{\frac12}\right|}
{\left|\sinh^2(\frac{\pi^2}{4R^2}-\frac{1}{2R})^{\frac12}\right|}
\label{oos-analytic}
\end{eqnarray}
One sees  the quadratic approximation 
predicts the explicit form of $C(G)$
and the fact  $\gamma=1$ in (\ref{asymptotic}).
It is remarkable that the correction is independent of $N$
in the limit.
Namely, it implies that  the estimates for $\omega$ are unchanged.

\begin{figure}
  \begin{center}
    \epsfxsize=200pt
    \leavevmode
    \epsfbox{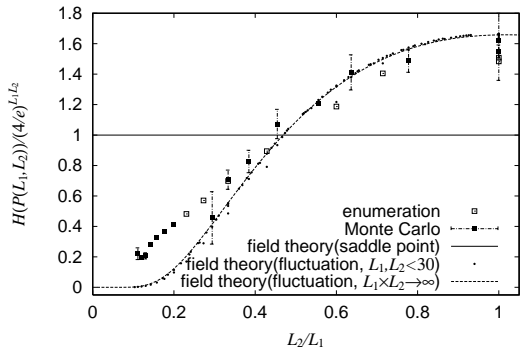}
  \end{center}
  \caption{
The number of Hamiltonian cycles for $P(L_1,L_2)$
for odd $L_j$ as a function of $L_2/L_1$.
Normalized by the saddle point result $(4/e)^{L_1L_2}$.
Plotted are the quadratic approximation to the field theory 
( solid square: $3\le L_1,L_2\le29$, 
dashed line: the limit $L_1\times L_2\rightarrow\infty$ ), 
exact results obtained by enumeration (box, $L_1\times L_2\le39$), and
estimates by weighted Monte Carlo simulations (solid box,
$L_1\times L_2\protect\lesssim90$). 
  \label{aspect.oop}
}
\end{figure}

\begin{figure}
  \begin{center}
    \epsfxsize=200pt
    \leavevmode
    \epsfbox{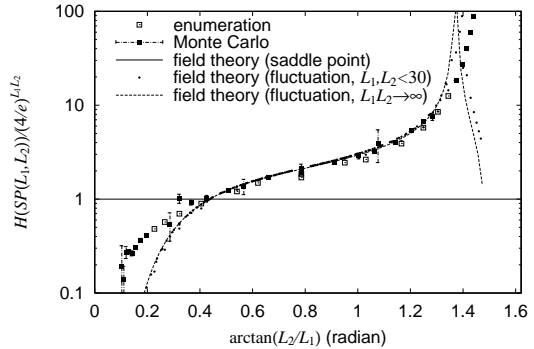}
  \end{center}
  \caption{
The log plot of the number of Hamiltonian cycles for $SP(L_1,L_2)$ for
odd $L_j$ as a function of $\arctan(L_2/L_1)$.
Normalized by the saddle point result $(4/e)^{L_1L_2}$.
Plotted are the same as in \protect Fig.\ref{aspect.oos}.
  \label{aspect.oos}
}
\end{figure}

Figs \ref{aspect.oop} and \ref{aspect.oos}
show the plot of (\ref{oop-analytic}), (\ref{oos-analytic}) normalized
by $(4/e)^{L_1L_2}$ 
as  functions of $L_2/L_1$  (solid line).
The finite $L_1L_2$ results (\ref{fluctuation_estimate})
for odd $L_j$ such that $3\le L_j \le 29$ are also shown (solid circle).
For the purpose of comparison, I plot the exact numbers of Hamiltonian
cycles (box) and estimates of them by simulations (solid box) for
$L_j$ odd\cite{even}.

The exact numbers are determined by 
direct enumeration by a program in C.
Table \ref{enumeration} shows the result.
This program is useful only for small graphs because the time needed
for calculation increases exponentially with $L_1 \times L_2$.
The simulation is based on the biased Monte Carlo method with a code 
in C.
The time needed grows again exponentially with $L_1 \times L_2$ but
with a smaller exponent.
I have used direct enumeration for $L_1\times L_2\le 39$ and Monte Carlo
simulation for  
$L_1 \times L_2 \lesssim 90$.
The evaluation of (\ref{fluctuation_estimate}) takes time proportional 
to $L_1\times L_2$. Therefore eq.(\ref{fluctuation_estimate}) has
an advantage even if it is approximate.

Figs \ref{aspect.oop} and \ref{aspect.oos} suggest that the
quadratic approximation to the field theory is reliable.
The field theory succeeds in predicting that 
the correction depends almost only on $L_2/L_1$ and that 
there is a qualitative difference between $P$ and $SP$.
The $L_2/L_1$-dependence of the correction generally
agrees with the exact result though there is a slight  deviation in
the small $L_2/L_1$ region. 

There is a definite discrepancy at $L_2/L_1=\pi^2/2$ for the
skew-periodic case. 
The quadratic approximation (\ref{oos-analytic}) diverges at
$L_2/L_1=\pi^2/2$ as seen in Fig.\ref{aspect.oos} while the exact
result takes finite values and is simply increasing there.
Actually, one can argue that $L_2/L_1\gtrsim\pi^2/2$ is out of range of
application of (\ref{fluctuation_estimate}). 
As I mentioned, I defined the limit $n\rightarrow0$ as the
continuation from $n>1$ where the gaussian integral converges.
Let us look at the evolution of the spectrum of 
$(A^{\rm{L}}(n)\Delta)$ on the way  from $n>0$ down to $n=0$.
The eigenvalue for $k_j=(1-1/L_j)\pi$ mode 
hits zero at some $0<n_{\rm c}<1$ if 
\begin{equation}
  2I_{L_1L_2,0} < \cos\frac{\pi}{L_1L_2}+\cos\frac{\pi}{L_2}.
  \label{additional_zero}
\end{equation}
This condition is equivalent to $L_2/L_1>\pi^2/2$ in 
the limit $L_1 \times L_2\rightarrow\infty$.
So one suddenly has a zero mode at $n=n_{\rm c}$ 
and no symmetry is responsible for it.
This suggests that the quadratic approximation breaks down there 
and that (\ref{oos-analytic}) is not reliable for such values of
$L_2/L_1$. A careful analysis shows that
the periodic case is free from such spurious zero modes.

\section{Summary and discussions}
I have found an approximate formula  for 
the number of Hamiltonian cycles of graphs 
by the quadratic approximation to the 
field theoretic representation.
It has been tested for 2d square lattice with a variety of boundary conditions.
I have obtained a multiplicative correction which is
independent of the size $N=\#V$.
Namely, the dependence on the boundary conditions is obtained and is
shown to be in a good agreement with the true behavior.

A natural extension to the present  problem is the  counting of 
the number of closed self-avoiding walks that 
visit $fN$ vertices of a graph, where $0\le f \le 1$ is fixed. 
In that case, the form (\ref{asymptotic}) can still be assumed but 
$C(G), \gamma$ and $\omega$ now depend  on $f$.
There is a field theoretic representation for 
$\omega(f)$ called lattice cluster theory\cite{Freed:lct}.
It is not identical with (\ref{field_theory_rep}) even for $f=1$.
It is interesting to compare the quadratic approximation to lattice cluster
theory\cite{NeCo:dilutedense,NeCo:packing} with the present result.
In the former one, estimate for $\omega(f)$ is improved over the 
saddle point approximation for $f<1$.
For $f=1$, the estimate for $\omega=\omega(1)$ is unchanged
in accord with the present case.

\section*{Acknowledgments}
I thank T.~Yamaguchi for collaboration in the initial stage of this
work. Many helpful conversations with S.~Hikami, M.~Itakura,
K.~Minakuchi, R. Pnini and J. Suzuki are gratefully acknowledged.  This work is
partially supported by Grant-in-Aid's for Scientific Research
(No. 08740196 and No. 08454106) from the Ministry of Education,
Science and Culture,  and by CREST from Japan Science and Technology
Corporation.


\begin{references}
\bibitem{invertibility}
In the case $\det \Delta=0$, 
the action (\ref{action}) is ill-defined.
In that case I can justify (\ref{field_theory_rep}) by  replacing
$\Delta$ with, for example,  
\begin{equation}
  \Delta(\epsilon_0)_{rr'} = \Delta_{rr'} + \epsilon_0 \delta_{rr'},
\end{equation}
where $\epsilon_0$ is a small number.
Then I have $\det(\Delta(\epsilon_0))\neq0$ and the field theory yields
\begin{equation}
  \langle \phi_j(r)\phi_k(r') \rangle = \delta_{jk} \Delta(\epsilon_0)_{rr'}
  \rightarrow \delta_{jk} \Delta_{rr'} \quad ( \epsilon_0 \rightarrow0)
\end{equation}
as desired. So eq.(\ref{field_theory_rep}) is  valid with
$\epsilon_0\rightarrow0$ limit understood.
By this regularization, the estimates
(\ref{mean_field_saddle_point_estimate}) 
and  (\ref{fluctuation_estimate}), which are free from
$\Delta^{-1}=\Delta(0)^{-1}$, remain valid.
\bibitem{even}
For the case with $L_j$ even, the $L_2/L_1$-dependence does not seem
to be the same. Work on this point is in progress.
\end{references}

\begin{table}[htbp]
\caption{Exact number of Hamiltonian cycles for $L_1\times L_2$ 2d
  square lattice with periodic ($P$) and skew-periodic ($SP$) boundary
  conditions.
  Determined by  the direct enumeration.\label{enumeration}}
\begin{tabular}{rrrrr}
$L_1$ & $L_2$ & $H(P(L_1,L_2))$ &$H(SP(L_1,L_2))$ &$H(SP(L_2,L_1))$ \\\hline
3& 3& 48& 55& 55\\
5& 3& 390& 397& 866\\
5& 5& 23580& 29001& 29001\\
7& 3& 2982& 2989& 13021\\
7& 5& 1045940& 1108006& 1820582\\
9& 3& 23646& 23653& 195157\\
11& 3& 196086& 196093& 2924373\\
13& 3& 1682382& 1682389& 43820323\\
\end{tabular}
\end{table}

\widetext
\end{document}